\documentclass[fleqn,usenatbib]{mnras}

\usepackage{newtxtext,newtxmath}

\usepackage[T1]{fontenc}
\usepackage{ae,aecompl}

\usepackage{graphicx}    
\usepackage{amsmath}    
\usepackage{bm}
\usepackage{natbib}
\usepackage{color,units}
\usepackage{float}
\usepackage[caption=false]{subfig}
\usepackage{aas_macros}
\usepackage{hyperref}

\def\swift{\textit{Swift}}
\newcommand{\bilby}{{\sc Bilby}}
\newcommand{\dynesty}{{\sc dynesty}}
\newcommand{\ciao}{{\sc ciao}}

\newcommand{\magnetarmodel}{\mathcal{M}_{\textrm{mag}}}

\newcommand{\radlossfull}{\mathcal{M}_{\textrm{rad-loss}}}

\title[Magnetar central engine with radiative losses]{Interpreting the X-ray afterglows of gamma-ray bursts with radiative losses and millisecond magnetars}

\author[N.Sarin et al.]{
Nikhil Sarin,$^{1,2}$\thanks{E-mail: nikhil.sarin@monash.edu}
Paul. D. Lasky$^{1,2}$
and Gregory Ashton$^{1,2}$
\\
$^{1}$School of Physics and Astronomy, Monash University, Vic 3800, Australia\\
$^{2}$OzGrav: The ARC Centre of Excellence for Gravitational Wave Discovery, Clayton VIC 3800, Australia}

\date{Accepted XXX. Received YYY; in original form ZZZ}

\pubyear{2020}

\begin{document}
\label{firstpage}
\pagerange{\pageref{firstpage}--\pageref{lastpage}}
\maketitle

\begin{abstract}
The spin-down energy of millisecond magnetars has been invoked to explain X-ray afterglow observations of a significant fraction of short and long gamma-ray bursts. 
Here, we extend models previously introduced in the literature, incorporating radiative losses with the spin down of a magnetar central engine through an arbitrary braking index. 
Combining this with a model for the tail of the prompt emission, we show that our model can better explain the data than millisecond-magnetar models without radiative losses or those that invoke spin down solely through vacuum dipole radiation. We find that our model predicts a subset of X-ray flares seen in some gamma-ray bursts. We can further explain the diversity of X-ray plateaus by altering the radiative efficiency and measure the braking index of newly-born millisecond magnetars.
We measure the braking index of GRB061121 as $n=4.85^{+0.11}_{-0.15}$ suggesting the millisecond-magnetar born in this gamma-ray burst spins down predominantly through gravitational-wave emission.
\end{abstract}

\section{Introduction}\label{sec:intro}
Cosmological gamma-ray bursts are the most energetic explosions in the Universe. They are historically split into two categories: long and short based primarily on their observed T$90$ duration, i.e., the duration where $90\%$ of the energy is released. 
Long gamma-ray bursts (T$90 \gtrsim \unit[2]{s}$) are typically associated with the collapse of massive stars and often accompanied by core-collapse supernovae such as the case for long gamma-ray burst GRB111209A and SN2011kl~\citep{grenier2015}. Short gamma-ray bursts (T$90 \lesssim \unit[2]{s}$) are associated with the merger of two compact objects such as a binary neutron star. 
The association of a binary neutron star merger with a short gamma-ray burst was confirmed by the coincident detection of short gamma-ray burst GRB170817A and gravitational waves from the binary neutron star inspiral GW170817~\citep{GW170817, abbott17_gw170817_gwgrb}.

Regardless of the progenitor, both long and short-duration gamma-ray bursts are accompanied by lower energy extended emission referred to as an afterglow. 
Traditionally, the origin of this afterglow has been attributed to the interaction of the relativistic outflow with the surrounding environment~\citep[e.g.,][]{Meszaros1993, Piran1998, Sari1998,  Meszaros1999, Zhang2006}. 
These external-shock fireball models have been largely successful in interpreting the afterglows of a large fraction of broadband afterglows of gamma-ray bursts.
However, more recently and in particular since the launch of the \textit{Neil Gehrels Swift Telescope}~\citep{swift}, X-ray afterglows of gamma-ray bursts have been observed in significantly more detail highlighting potential problems for the external-shock models.
In particular, two observed features of X-ray afterglows are problematic to explain with the fireball model; the extended plateau seen in $\approx 50\%$~\citep[e.g.,][]{Rowlinson2013} of gamma-ray burst afterglows and the sharp drop in luminosity seen in $\approx 20\%$~\citep[e.g.,][]{gao16}. 
These observational features are well interpreted within the framework of additional energy injection from a rapidly-spinning, highly magnetic neutron star, referred to as a millisecond magnetar.
Determining whether the central engine is a black hole or a neutron star has important implications for the nuclear equation of state, the progenitors and rates for fast radio bursts, and the jet-launching mechanism for gamma-ray bursts (see \cite{kumar15} and references therein).

Millisecond magnetars were first proposed by \citet{usov92, Dai1998, Zhang2001} as a central engine for gamma-ray bursts.
The millisecond-magnetars spin-down energy provides an additional energy source that powers the X-ray afterglow. 
Such a model has been broadly successful in explaining the two aforementioned observational features~\citep[e.g.,][]{fan06, Rowlinson2010, Rowlinson2013, DallOsso2011, Lu2015, LaskyLeris2017}; the plateau, which is sustained through the additional energy injection, and the sharp drop in luminosity, which is attributed to the collapse of these rapidly-spinning neutron stars into black holes~\citep[e.g.,][]{Rowlinson2010, Lasky2014, sarin20}. 

Although broadly successful in explaining these two features, the millisecond-magnetar model fails to explain other observations. 
For example, the magnetar model can only explain the X-ray afterglow and has no detailed prescription for emission in other electromagnetic bands which is instead attributed to the external shock~\citep[e.g.,][]{DallOsso2011}.
This emission from the external shock is believed to be subdominant in the X-ray afterglow when a millisecond magnetar is active.
This seems plausible as, in a subset of short gamma-ray bursts that have observations of a sharp drop, one can see the previously subdominant emission from the external shock again~\citep[e.g.,][]{Rowlinson2013, sarin20}.
Ultimately, a complete model is needed which predicts the emission across the electromagnetic spectrum.
Work by \citet{metzger14, strang19} towards this goal assume that the energy from the spin down of the millisecond magnetar is dissipated through a wind, similar to a pulsar-wind nebula. However, such models have not been fit in detail to observations.

The spin down of a magnetar can be characterised by its braking index. Early efforts in modelling the X-ray afterglow with the magnetar model involved assuming the magnetar was spinning down solely through magnetic dipole radiation~\citep{Zhang2001,fan06, Rowlinson2013, Lu2015}. 
This assumption is in contrast to observations that suggest newly-born magnetars spin down through the emission of gravitational waves~\citep{fan13,gao16,sarin20}.
Under the assumption that the braking index is arbitrary but constant through time,~\citet{LaskyLeris2017} measured the braking index of two gamma-ray bursts. 
More recently,~\cite{mus19} measured the evolution of the braking index through the coupling of the braking index to the evolution of the magnetic inclination angle~\citep[see e.g.,][]{lander18}. 
However, perhaps more critically, these works assume a constant efficiency in converting between the spin-down energy of the magnetar central engine and the resulting X-ray afterglow luminosity, assuming that $\sim 10\%$ of the central engine spin-down energy is converted into an X-ray luminosity.
Given the diversity of gamma-ray burst afterglows and their environments, it is difficult to conceive of all systems behaving in the same way through time and with the same efficiency.

The idea of a non-constant and/or distinct efficiency has been explored previously. 
\citet{xiao19} model the efficiency as dependent on the luminosity of the central engine itself i.e., $\eta_{\text{x-ray}} \propto L_{\text{magnetar}}$, where $\eta_{\text{x-ray}}$ is the efficiency and $L_{\text{magnetar}}$ is the luminosity of the magnetar.
This suggests that during the plateau phase, the efficiency stays constant as the luminosity of the millisecond magnetar is roughly constant, while at late times the efficiency drops following the drop in luminosity from the central engine.
Another approach to account for efficiency is by considering the effect of radiative losses due to the deceleration of the shock in the interstellar medium~\citep{cohen98,cohen99}.
\citet{DallOsso2011} developed such a model where they considered the effect of radiative losses for a millisecond magnetar spinning down solely through vacuum dipole radiation, a model that has since been fit to several gamma-ray burst afterglows assuming the magnetar emission has an angular structure~\citep{Stratta2018}.

Here, we extend the model from \citet{DallOsso2011} by including spin down through an arbitrary braking index and by incorporating the emission from the tail of the prompt. 
We fit our model to a sample of well-studied long and short gamma-ray bursts that have been previously suggested to have millisecond magnetar central engines.
We find that our model can explain some X-ray flares seen in the X-ray afterglow of some gamma-ray bursts, and is a better fit to the data than millisecond magnetar models used currently in the literature. 
In the process, we also measure the braking index of these millisecond magnetars.
We introduce our model for a millisecond-magnetar spinning down through arbitrary braking indices and including radiative losses in Sec. \ref{sec:model}.
We then present our results for a small subset of long and short gamma-ray bursts in Sec. \ref{sec:results}.
We discuss the implications of our results and conclude in Sec. \ref{sec:implications} and \ref{sec:conclusion} respectively.
\section{Model}\label{sec:model}
We model the emission in the X-ray afterglow of the gamma-ray burst to be a combination of energy injection from a newly born millisecond magnetar interacting with the surrounding environment resulting in radiative loss~\citep{cohen99} and incorporating the emission from the tail of the prompt.
As mentioned in Sec.~\ref{sec:intro}, such a model without the inclusion of the emission from the tail of the prompt and assuming the magnetar spins down solely through vacuum dipole radiation was introduced by~\cite{DallOsso2011}.
Our extension to this model starts by generalising the spin down of the magnetar through an arbitrary braking index such that $\dot{\Omega} \propto \Omega^{n}$. Here, $\Omega$ and $\dot{\Omega}$ are the neutron stars angular frequency and its derivative respectively, and $n$ is the braking index. 
This generalisation allows one to write the luminosity of a millisecond magnetar spinning down through an arbitrary braking index~\citep{LaskyLeris2017},
\begin{equation}\label{eqn:magnetar}
L_{\mathrm{sd}}(t) = L_{0}\left(1 + \frac{t}{\tau}\right)^{\frac{1 + n}{1 - n}}.
\end{equation}
Here, $L_{\mathrm{sd}}$ is the spin-down luminosity of the magnetar, $t$ is the time since burst, and $\tau$ is the spin-down timescale.
The spin-down energy of the magnetar is subject to some radiative loss at the shock interface, which implies~\citep{DallOsso2011},
\begin{equation}\label{eq:energybalance}
\frac{dE}{dt} = L_{\mathrm{sd}} - \kappa\frac{E}{t}.
\end{equation}
Here, 
\begin{equation}{\label{eq:kappa}}
\kappa = 4\epsilon_{e}\frac{d\ln t^*}{d\ln t},
\end{equation}
is the radiative efficiency, $\epsilon_e$ is the fraction of total energy transferred to the electrons, and $d\ln t^{*}/d\ln t$ describes the dynamical evolution of the shock where $t^*$ is the time in the reference frame of the central engine where the energy is transferred into the shock.
In Eq.~(\ref{eq:energybalance}), the first term on the right-hand side captures the energy injection from the spin down of the neutron star central engine, while the second term captures radiative losses at the shock interface.
The lightcurve as seen by a distant observer is then, 
\begin{equation}\label{eq:radiative_losses}
L(t) = At^{\Gamma} + \mathcal{H}(t - t_{0})\kappa\frac{E(t, t_0)}{t}.
\end{equation}
Here, $E(t, t_0)$ is the solution to Eq.~(\ref{eq:energybalance}), $t_0$ is the time at which the observer starts to see the emission from radiative losses, $A$ and $\Gamma$ are the power-law amplitude and power-law exponent, respectively, which together describe the emission from the tail of the prompt. A lower limit on $t_0$ is the afterglow onset time, i.e., the time it takes the blast wave to reach the deceleration radius~\citep[e.g.,][]{Sari1998, Sari1999}. 

The tail of the prompt emission is the power-law decay in flux associated with the curvature effect. Photons emitted at the same time but at different latitudes within the jet opening angle will arrive at the distant observer at different times due to propagation effects, resulting in a steep temporal decay~\citep[e.g.,][]{kumar2000, Zhang2005}. 
The tail of the prompt therefore marks the transition from the prompt emission phase to the afterglow emission. 
Furthermore, given typical X-ray afterglows do not show an early rise, the afterglow onset time and the associated rise in flux is likely hidden by the emission from the tail of the prompt.
We note that previous works involving radiative losses did not include the tail of the prompt emission in their fit to minimise fitting parameters~\citep{DallOsso2011, Stratta2018}. 
In later sections, we show that the inclusion of the tail of the prompt and radiative loss subject to energy injection from a newly-born neutron star can explain several interesting aspects of gamma-ray burst X-ray afterglows.
\section{Results}\label{sec:results}
We fit our model (Eq.~\ref{eq:radiative_losses}) to the X-ray afterglow of a small sample of short and long-duration gamma-ray bursts observed by \swift{} using the nested sampler \dynesty~\citep{dynesty} through the Bayesian inference library \bilby~\citep{bilby} and a Gaussian likelihood.
Our selection of gamma-ray bursts are chosen as their X-ray afterglow has a shallow decay phase indicative of central engine activity. 

For our sample of gamma-ray bursts, we use the $0.3-10~\unit{keV}$~flux from the \swift{} database using the automatic binning strategies~\citep{evans09,evans10}.
We convert the flux into luminosity using \ciao~\citep{ciao} performing k-corrections~\citep[e.g.,][]{Bloom_2001}. 
The gamma-ray bursts analysed, their associated redshifts and T$90$ durations are summarised in Table~\ref{table:GRBs_analysed}.
For gamma-ray bursts without a measured redshift, we assume a fiducial redshift $z = 0.75$ so that our model can be fit to luminosity data.
\begin{table}
\centering
 \caption{Gamma-ray bursts analysed along with their associated T$90$ duration and redshift.}
 \label{table:GRBs_analysed}
 \begin{tabular}{ccc}
  \hline
GRB & T90(s) & Redshift \\
  \hline
GRB050319 & $152.5$ & $3.24$\\
GRB051221A & $1.4$ & $0.547$\\
GRB060313 & $0.7$ & N/A\\
GRB060729 & $115.3$ & $0.54$\\
GRB061121 & $81.3$ & $1.314$\\
GRB070809 & $1.3$ & $0.2187$\\
GRB080430 & $16.2$ & $0.767$\\
GRB111020A & $0.4$ & N/A\\
  \hline
 \end{tabular}
\end{table}

By including the effect of spindown through an arbitrary braking index we have introduced a new model for explaining X-ray afterglows of gamma-ray bursts.
However, a pertinent question to consider: is the data better explained by the model? 
We answer this question through Bayesian model selection following the procedure in \cite{sarin19}.
We perform model selection for two models: a millisecond-magnetar model with an arbitrary braking index (Eq.~\ref{eqn:magnetar}) and the radiative losses model introduced here.

Our priors for the different models are listed in Table~\ref{table:priors}. 
We note that we used the same priors for all gamma-ray bursts except GRB051221A and GRB070809 which both have a narrower prior on $t_0$ to ensure the sampler converges to the correct mode. This tighter prior choice implies that the effect of radiative losses, and by extension the afterglow onset, occurs earlier in these short gamma-ray bursts. 
In reality, $t_0$ should be informed by considering the spectra of the gamma-ray burst itself. The transition from the tail of the prompt to the afterglow will be marked by a spectral change which then provides a tight constraint on $t_0$.  However, given the difficulty in identifying a spectral change in gamma-ray burst data and the additional fitting required we use a more agnostic prior. 
\begin{table}
\centering
 \caption{Priors for the radiative losses model with spindown through an arbitrary braking index ($\mathcal{M}_{\textrm{rad-loss}}$). The priors for the millisecond magnetar model ($\mathcal{M}_{\textrm{mag}}$) are identical except for $\kappa$, $t_0$ and $\log_{10}E_0$ which are parameters not applicable to this model. We note that a LogUniform prior is a prior that is uniform in log-space.}
 \label{table:priors}
 \begin{tabular}{cc}
  \hline
  Parameter [Units]& $\radlossfull$ \\ \hline
$A [\unit[10^{50}]{erg}]$ &  $\textrm{LogUniform}[10^{-10},10^{15}]$\\
$\Gamma$& $\textrm{Uniform}[-7,1]$\\
$L_{0}[\unit[10^{50}]{erg}]$ & $\textrm{LogUniform}[10^{-5},1]$\\
$\tau$ [\unit{s}]&$\textrm{LogUniform}[10^{2},10^{7}]$\\
$n$ & $\textrm{Uniform[1.1,7]}$\\
$\kappa$ &$\textrm{LogUniform[$10^{-3}$,4]}$\\
$t_{0}$ [\unit{s}]& $\textrm{Uniform}[30,400]$\\
$\log_{10}E_0$ & $\textrm{Uniform}[-10,2]$\\
  \hline
 \end{tabular}
\end{table}

The Bayes factors\footnote{For clarity, we note that $BF_{a/b} = 2$ indicates model $a$ is twice as likely as model $b$.} for our analysis are shown in Table~\ref{table:bayesfactors}. Typically, a Bayes factor $\gtrsim 100$ is considered to be decisive~\citep{krass95}.
The corner plots showing the one and two-dimensional posterior distributions for all gamma-ray bursts are available online~\citep{git_repository}.
\begin{table}
\centering
 \caption{Gamma-ray bursts analysed along with the $\ln BF$ for the radiative losses~(Eq.~\ref{eq:radiative_losses}) model compared with the magnetar model~(Eq.~\ref{eqn:magnetar})}
 \label{table:bayesfactors}
 \begin{tabular}{cc}
  \hline
  GRB & $\ln BF_{\radlossfull/\magnetarmodel}$ \\
  \hline
GRB050319 & $3.1$ \\
GRB051221A & $160.2$ \\
GRB060313 & $183.7$ \\
GRB060729 & $141.2$ \\
GRB061121 & $241.2$ \\
GRB070809 &  $0.3$ \\
GRB080430 & $51.4$ \\
GRB111020A & $93.9$\\
  \hline
 \end{tabular}
\end{table}
We find that all eight gamma-ray bursts analysed favour the inclusion of radiative losses over the magnetar model. The weakest support comes from GRB070809 which has a weak preference for the radiative losses model. In other words, for this gamma-ray burst, the inclusion of the additional radiative losses physics does not provide a significantly better fit to the data. This weak preference may indicate that the effect of radiative losses is negligible in this gamma-ray burst or that more simply, there is insufficient data to probe the effects of this model. We return to this point in Sec~\ref{sec:implications}.  
\subsection{X-ray flares}\label{subsec:flares}
Flares are fast-rising then exponentially decaying features seen in several long and short gamma-ray bursts. 
While more prevalent in long-duration gamma-ray bursts, they have been observed in several short gamma-ray bursts as well, suggesting the mechanism behind them may be universal \citep[e.g.,][]{perna06}. 
However, they are also diverse and no one mechanism can successfully interpret the different characteristics~\citep[e.g.,][]{kumar15}.

A subset of flares are seen at the onset of the X-ray afterglow of a large fraction of gamma-ray bursts~\citep{OBrien_2006}.
Here, the onset of the afterglow marks the transition from the steep decay attributed to the tail of the prompt emission. A flare near this transition is difficult to explain with an external shock origin, and has been suggested to require central engine activity~\citep{Zhang2005}, or specifically in the case of short gamma-ray bursts, magnetic reconnection events~\citep{Fan_2005b}. 

We find that our model can explain these flares as the breakout of excess energy in the relativistic blast wave at the onset of the afterglow. 
Here specifically, the flare is the product of the excess energy and transition to emission described by radiative losses with a millisecond magnetar central engine (i.e., the transition to the second term on the right-hand side in Eq.~\ref{eq:radiative_losses}).
The size of the flare is related to the amount of energy that is in the relativistic blast wave at the onset of radiative losses. 
The decay indices of the flare itself are determined by the radiative efficiency $\kappa$; in general, smaller $\kappa$ produce more gradually decaying flares. 
Although this mechanism can successfully explain the diversity in size and decay index of flares seen in gamma-ray bursts, it likely cannot explain multiple flaring episodes. 
The excess energy will likely only generate one flare and such a flare will occur at the onset of the afterglow emission implying that other flares must be generated differently. 
Flares that occur later in the X-ray afterglow may also be products of radiative losses and excess energy but to explain such features the energy injection mechanism will need to be modified from the model we have used (Eq.~\ref{eqn:magnetar}).
\begin{figure}
\centering
\includegraphics[width=0.5\textwidth]{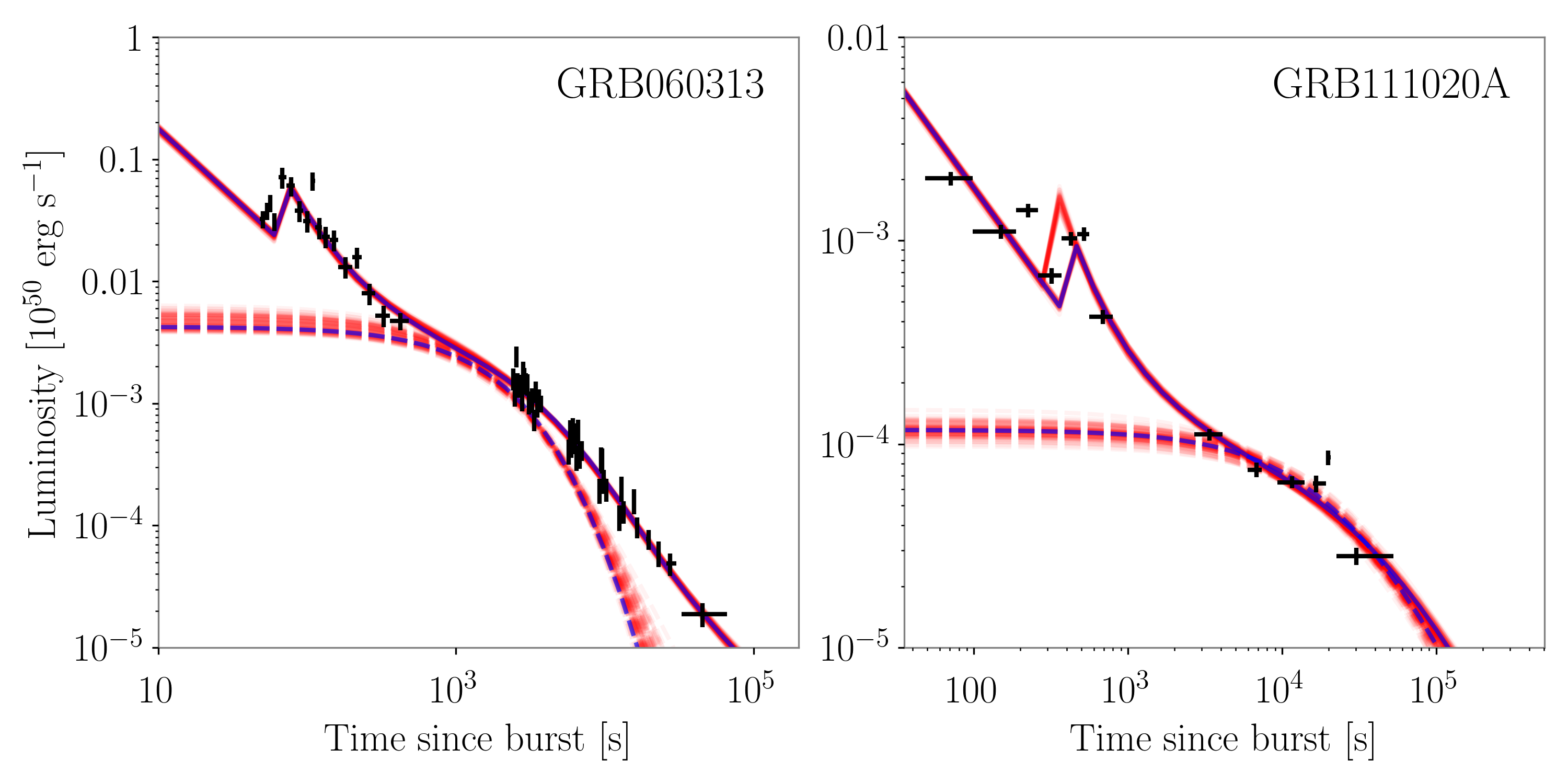}
\caption{X-ray lightcurves for two short gamma-ray bursts. Black points indicate data from \textit{Swift}. The blue curve shows the maximum likelihood model for the radiative losses model (Eq.~\ref{eq:radiative_losses}). The dark red band is the superposition of $100$ models randomly drawn from the posterior distribution. The flare seen in the onset of the plateau phase is naturally explained by the radiative losses model. We also show the underlying spin-down luminosity from the nascent magnetar in dashed lines.}
\label{fig.flares}
\end{figure}

In Figure~\ref{fig.flares}, we show our fit to two short gamma-ray bursts, GRB060313 and GRB111020A which have flares near the transition of the tail of the prompt and the afterglow. Our model successfully explains the flare size and decay while also being a good fit for the rest of the data. In particular, GRB111020A has a bi-modality in the location of the flare. This is a product of the uncertainty in $t_{0}$ (i.e., the time where radiative losses turn on) given the sparsity of the data near the flare this parameter is poorly constrained, resulting in a bi-modality in when the flare occurs. Given the magnetar model without radiative losses (Eq.~\ref{eqn:magnetar}) cannot explain flares, it is not surprising that both these gamma-ray bursts strongly favour the radiative losses model (see Table~\ref{table:bayesfactors}).
\subsection{Long gamma-ray bursts}\label{subsec:lgrb}
Long gamma-ray bursts are associated with the collapse of massive stars. The afterglow of these bursts has been extensively studied, and for the vast majority of gamma-ray bursts, been largely in agreement with the predictions of the external shock model.
A few gamma-ray bursts do, however, have sharp drops or plateaus indicative of a magnetar central engine~\citep[e.g.,][]{troja07,lyons10, beniamini17b}, in particular, GRB050319, GRB060729, GRB061121, and GRB080430~\citep[e.g.,][]{DallOsso2011, xiao19, lu19}.
These four gamma-ray bursts are well studied, partly due to their plentiful observations and have been fitted with the millisecond-magnetar model on numerous occasions~\citep[e.g.,][]{DallOsso2011, xiao19}. 
Notably, the former included the effect of radiative losses with vacuum dipole radiation, while the latter assumed the X-ray luminosity is entirely from vacuum dipole radiation but the magnetar was spinning down through an arbitrary braking index.

\begin{figure}
        \includegraphics[width=0.5\textwidth]{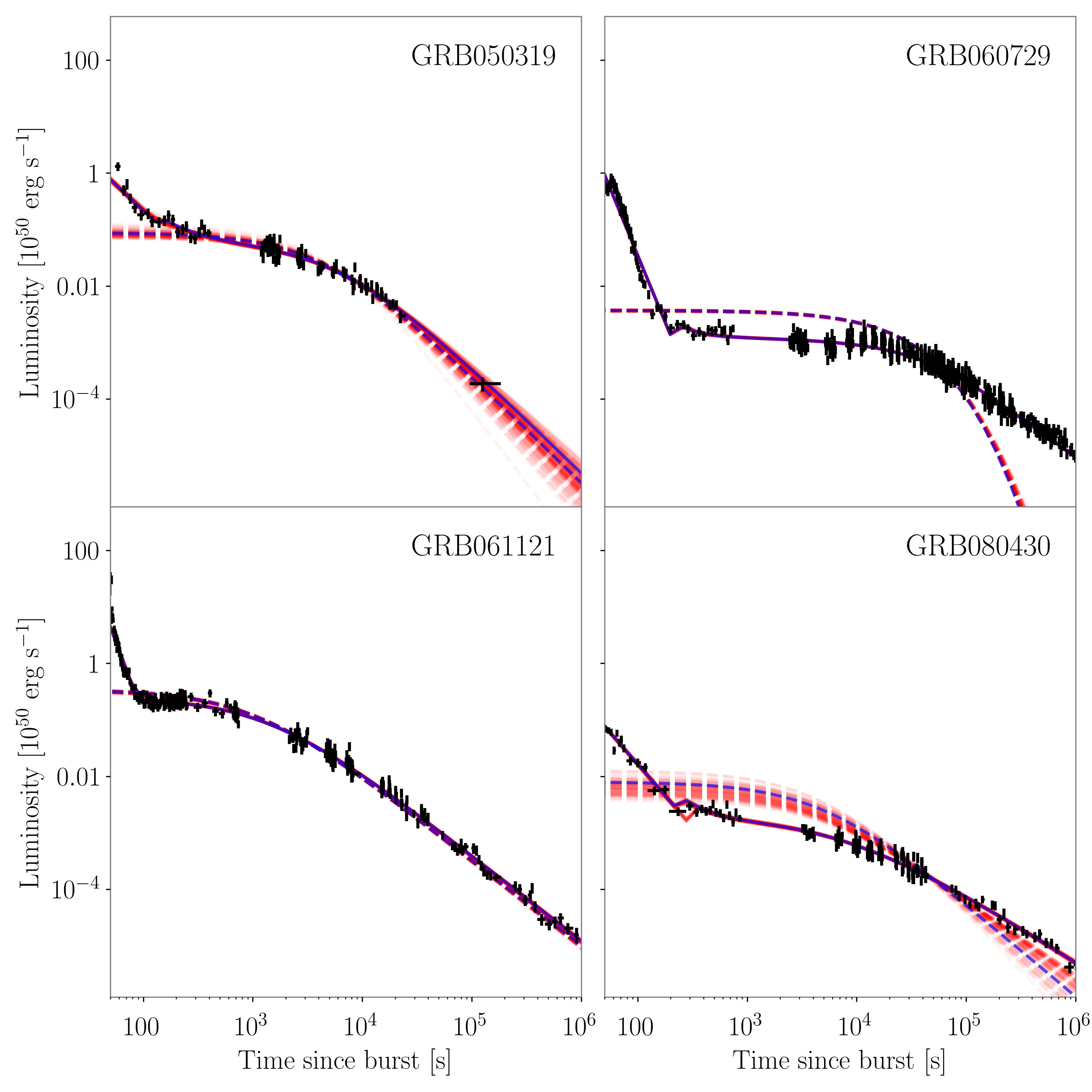} \hspace{-.4cm}
  \caption{X-ray lightcurves for four long gamma-ray bursts. Black points indicate data from \textit{Swift}. The blue curve shows the maximum likelihood model for the radiative losses model (Eq.~\ref{eq:radiative_losses}). The dark red band is the superposition of $100$ models randomly drawn from the posterior distribution. We also show the underlying spin-down luminosity from the nascent magnetar in dashed lines. For GRB050319 and GRB0601121 the observed emission closely follows the spin-down luminosity of the nascent magnetar, while for GRB060729 and GRB080430 the observed luminosity is significantly different. This is direct consequence of the different radiative efficiency $\kappa$ for these gamma-ray bursts.}
\label{fig.longs}
\end{figure}

We fit our model to these four aforementioned gamma-ray bursts, with our results shown in Figure~\ref{fig.longs}.
Since these gamma-ray bursts have plentiful observations, we are also able to constrain the inherent emission from the millisecond magnetar itself, which is shown as the dashed curves in Figure~\ref{fig.longs}. 
We note that the inherent emission of the millisecond magnetar for GRB050319 and GRB061121 closely follows the observed lightcurve suggesting the impact of radiative losses is minimal.
By contrast, GRB060729 and GRB080430 show vast differences between the observed lightcurve and the inherent emission from the magnetar, suggesting radiative losses play a critical role.
This impact of radiative losses is determined through $\kappa$, the radiative efficiency parameter, with lower values indicating radiative losses is more impactful.
Why the impact of radiative losses is different in these gamma-ray bursts is an intriguing question, which we discuss in more detail in Sec~\ref{sec:implications}.
These gamma-ray bursts are all well fit by the radiative losses model and comparing Bayes factors~(see~Table~\ref{table:bayesfactors}), they strongly favour the inclusion of radiative losses over the magnetar model. 
\subsection{Short gamma-ray bursts}\label{subsec:sgrb}
Short gamma-ray bursts are associated with the merger of compact objects.
The multimessenger observations of GW170817 confirmed that binary neutron star mergers are the progenitors of some short gamma-ray bursts~\citep{abbott17_gw170817_gwgrb, GW170817A_GRB}.
One of the motivations for determining whether millisecond magnetars exist in the aftermath of a short gamma-ray burst is to determine the maximum mass of neutron stars, and therefore the nuclear equation of state. 

Unlike long gamma-ray bursts, short gamma-ray bursts from neutron star mergers have a well-defined progenitor mass distribution, motivated by the galactic double neutron star distribution~\citep{Kiziltan2013}.
However, the recent detection of GW190425 suggests the local binary neutron star distribution may be a poor representation of binary neutron stars mergers~\citep{abbott19_190425_detection}.
Determining whether a short gamma-ray burst produced a black hole remnant or a millisecond magnetar can immediately inform the maximum mass. 
In reality, this is much more complicated as unless accompanied by gravitational waves from the inspiral, short gamma-ray bursts cannot alone provide a measurement for the maximum mass.
For GW170817, the only coincident binary neutron star merger and short gamma-ray burst to date (GW190425 did not have any coincident electromagnetic observation~\citep[e.g.,][]{Coughlin_2019, Hosseinzadeh_2019}), there is still no strong consensus on the fate of the post-merger remnant (see \cite{ai19} for a review for the different possibilities). 

\begin{figure}
\centering
\includegraphics[width=0.5\textwidth]{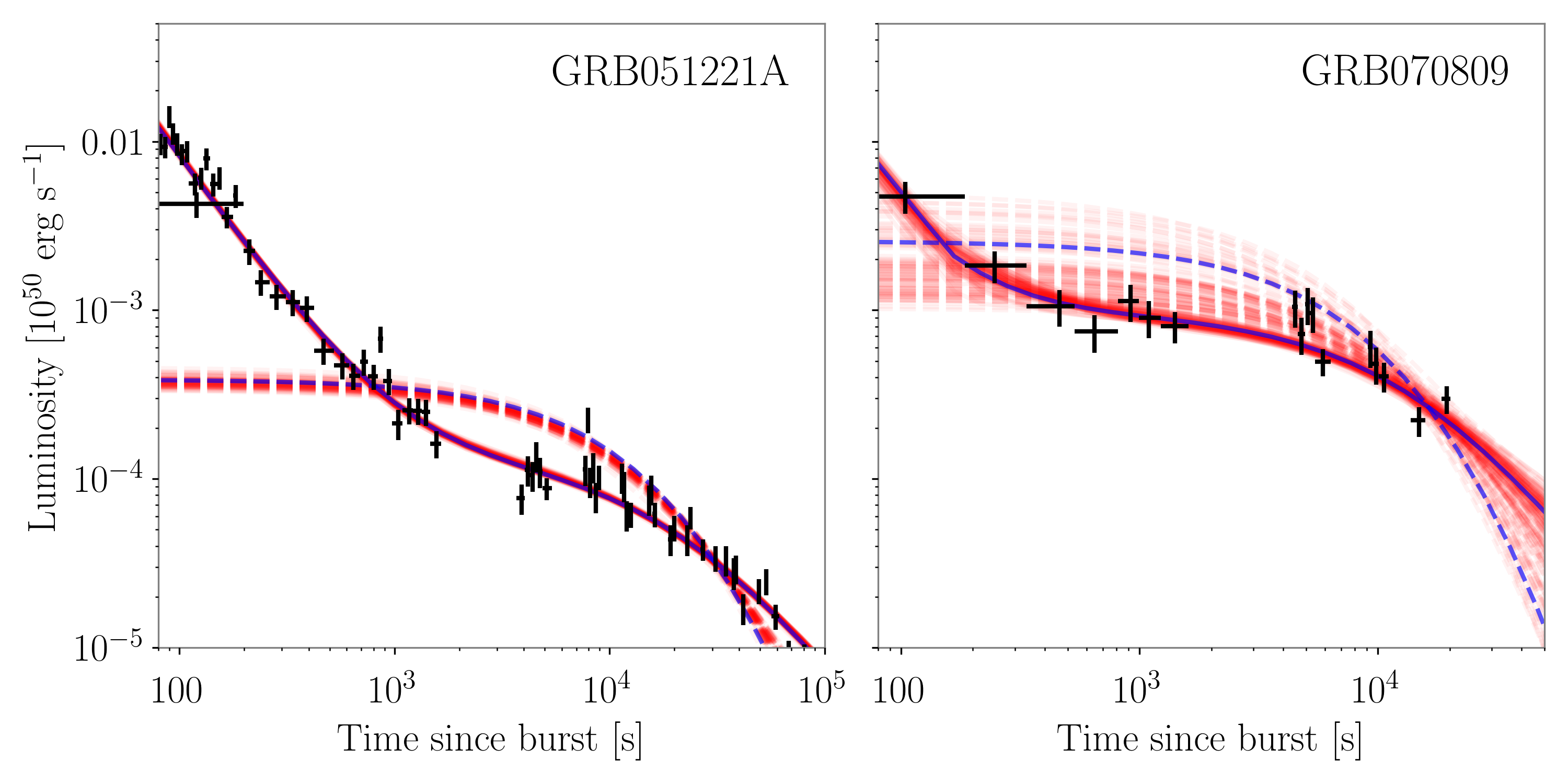}
\caption{X-ray lightcurves for four long gamma-ray bursts. Black points indicate data from \textit{Swift}. The blue curve shows the maximum likelihood model for the radiative losses model (Eq.~\ref{eq:radiative_losses}). The dark red band is the superposition of $100$ models randomly drawn from the posterior distribution. In dashed lines we show the underlying spin-down luminosity from the nascent magnetar.}
\label{fig.shorts}
\end{figure}

We use our model to analyse the afterglow of two short gamma-ray bursts: GRB051221A and GRB070809. 
The former is a well-studied gamma-ray burst commonly associated with a millisecond magnetar central engine~\citep{fan06, soderberg06}. However, it has been subject to significant debate with analysis by~\cite{Lu2015} finding the afterglow to have a post-jet break index $\alpha \approx -1$ which is consistent with an external shock model or suggestive of magnetar spin down through gravitational-wave emission. We discuss this point in greater detail in Sec.~\ref{subsec:brakingindex}.
GRB070809 is another short gamma-ray burst with a plateau in the X-ray afterglow suggestive of a neutron star central engine. Furthermore, it was recently identified to be associated with a blue kilonova counterpart~\citep{jin_070809} which naturally suggests a long-lasting neutron star central engine~\citep[e.g.,][]{Margalit2017}.
We find that our model can successfully explain the observations of both gamma-ray bursts with our fits shown in Figure~\ref{fig.shorts}.

Comparing Bayes factors for both the model with radiative losses and without, we see that GRB051221A strongly favours the inclusion of radiative losses. Furthermore, while the observed lightcurve is consistent with a post-jet break index of $\alpha \approx -1$, the inherent emission from the millisecond magnetar is significantly different, implying a different braking index. We discuss this in more detail in Sec.~\ref{subsec:brakingindex}.
GRB070809 has a weak preference for the model including radiative losses. 
This may be indicative of the small effect of radiative losses for this gamma-ray burst, but given the relatively small amount of data, it is equally likely that the data cannot distinguish between the two models significantly. This is apparent when looking at the inherent emission from the millisecond magnetar for GRB070809. 
\subsection{Braking index}\label{subsec:brakingindex}
As discussed in Sec.~\ref{sec:intro}, millisecond-magnetar models initially assumed the magnetar was spinning down solely through vacuum dipole radiation. This assumption was relaxed and used to measure the braking index of two millisecond magnetars born in GRB130603B and GRB140903A finding only the former to be consistent with $n=3$ associated with vacuum dipole radiation~\citep{LaskyLeris2017}. 
Newly born millisecond magnetars are not expected to spin down solely through vacuum dipole radiation, instead, implying a significant amount of early gravitational-wave emission~\citep[e.g.,][]{fan13, gao16, sarin20}.
Furthermore, mechanisms such as twisted magnetosphere~\citep[e.g.,][]{thompson02}, magnetic field axis evolution~\citep[e.g.,][]{Cutler2002} which are expected to be important in newly born millisecond magnetars~\citep[e.g.,][]{LaskyLeris2017,lander18} all predict the braking index $n\lesssim3$. 
Several more braking index measurements from gamma-ray bursts with putative millisecond magnetar central engines have been made~\citep{xiao19, lu19}, however, none of these consider the effect of radiative losses. 

\begin{figure}
\centering
\includegraphics[width=0.5\textwidth]{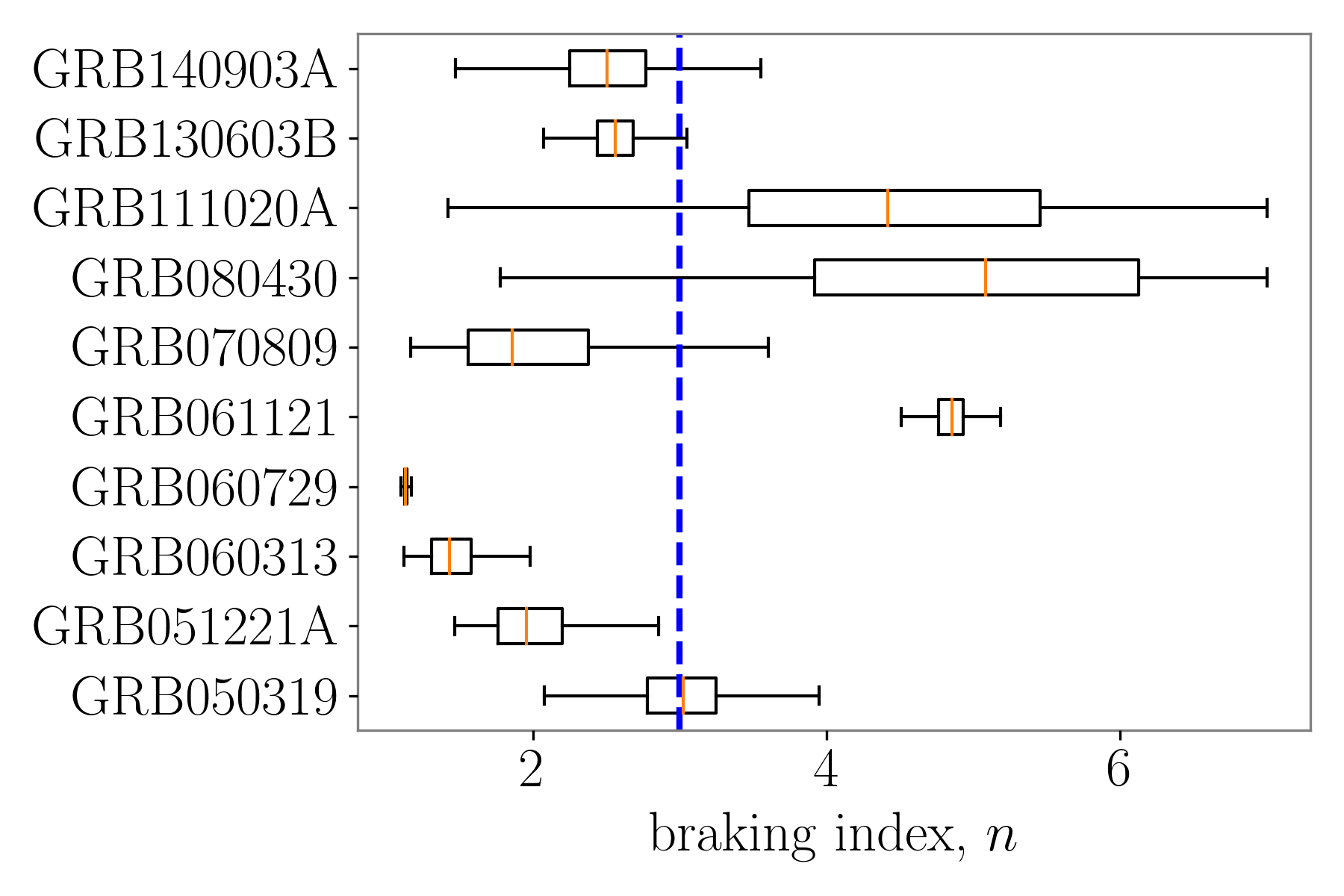}
\caption{Measured millisecond magnetar braking indices for all gamma-ray bursts analysed in this paper and GRB140903A and GRB130603B with the radiative losses model (Eq.~\ref{eq:radiative_losses}). The blue dashed line indicates $n=3$, the braking index associated with vacuum dipole radiation.}
\label{fig.brakingindex}
\end{figure}

In Fig~\ref{fig.brakingindex}, we show the braking index measurements with the radiative losses model for the eight gamma-ray bursts analysed in this paper. We also show the braking index measurement for the two aforementioned gamma-ray bursts, GRB130603B and GRB140903A which were measured previously~\citep{LaskyLeris2017} but we revisit with the radiative losses model.
In a simplistic view, the impact of radiative losses is to lower the braking index.
The braking index is measured by the slope of the curve after the plateau phase ends, a shallower slope indicating a higher braking index.
Inclusion of radiative losses means the shallower observations of the lightcurve can instead be explained by the radiative losses and therefore implies a steeper slope for the braking index.

For GRB051221A in particular, without the inclusion of radiative losses, we measure the braking index $n=4.51^{+0.45}_{-0.38}$. 
Such a high braking index would imply a significant amount of energy released in gravitational waves at up to $\sim10^5$~s post-formation. 
While not impossible, this is difficult to explain~\citep[e.g.,][]{Lasky2016}.
However, by including radiative losses, which is the preferred model, we measure $n=1.96^{+0.38}_{-0.27}$ ($68\%$ confidence interval), alleviating this concern.

The braking index of GRB061121 is also intriguing, we measure $n=4.85^{+0.11}_{-0.15}$ ($68\%$ confidence interval) with the radiative losses model which is consistent with the conclusion that the millisecond magnetar is spinning down predominantly through gravitational-wave emission. 
We do however caution that since this is a long gamma-ray burst, there may be additional effects, such as fall back accretion that may make such a measurement unreliable. 
At a redshift of $z=1.314$, the gravitational-wave emission from such an object will not be observable individually in aLIGO or with third-generation telescopes such as the Einstein Telescope~\citep[e.g.,][]{Sarin2018}.
However, it does suggest that millisecond magnetars born in long gamma-ray bursts may spin down through gravitational-wave emission and that such a population of gravitational-wave sources may be observable as part of the stochastic gravitational-wave background.
\section{Implications}\label{sec:implications}
The inclusion of radiative losses, the tail of the prompt, and the spin down through an arbitrary braking index can successfully explain several aspects of gamma-ray burst X-ray afterglows.
The radiative efficiency $\kappa$ controls the shape of the plateau and how much of the inherent emission from the central engine is directly visible to the observer. 
Higher values of $\kappa$ imply the observed lightcurve closely follows the inherent emission from the millisecond magnetar, while smaller values of $\kappa$ imply the effect of radiative losses is larger, and the observed lightcurve is visibly different from the emission from the millisecond magnetar.
In general, we notice that gamma-ray bursts in a host galaxy with a higher density have smaller $\kappa$ i.e., the impact of radiative losses is larger. 
This seems plausible as a denser medium likely means more radiative loss at the shock-interface. However, we leave the exploration of this correlation to future work with a larger selection of gamma-ray bursts.

Radiative losses can also explain the diversity in size and decay of X-ray flares seen at the onset of the afterglow. We have shown this for two gamma-ray bursts, GRB060313 and GRB111020A. In our model, the flare is a natural product of excess energy in the relativistic blast wave at the onset of the afterglow phase. Such a mechanism can only generate one flare, but we note that later flares may also be a product of radiative losses. However, modelling this will require a modification to the energy injection term we have used in this work.

In our model, $\kappa$ encodes two terms; $\epsilon_{e}$; the fraction of total energy transferred to electrons, and $d\ln t^{*}/d\ln t$, which describes the dynamical evolution of the shock.
The dynamical evolution of the shock is difficult to constrain and requires detailed hydrodynamical modelling which would not be sufficiently fast making the fitting procedure computationally difficult.
The former term is easier to probe, the afterglow emission from the external shock of a gamma-ray burst can provide a measurement for $\epsilon_{e}$. 
Unfortunately, one cannot use the X-ray observations to make this measurement as owing to the putative magnetar, the external-shock emission is likely subdominant, and if not, it is difficult to decouple the emission from the central engine and one from the external shock.
This motivates the need for a general model which includes the effect of both a millisecond magnetar and an external shock which we leave for future work.
If one could measure $\epsilon_{e}$ independently, through the afterglow observation in another electromagnetic band, for example~\citep{beniamini17a}, this would allow the decoupling of the two terms in $\kappa$ and direct measurement of the dynamical evolution of the shock. Under simple assumptions this could lead to a measurement of the decay index for the Lorentz factor and provide a complementary way of determining the structure of the jet. In this paper, we work only with the X-ray afterglow data and therefore cannot decouple the two parameters.

The radiative losses model introduced here can explain all the resolvable features in all eight gamma-ray bursts we have analysed. However, successfully fitting this model for all observed gamma-ray bursts is problematic. In particular, measuring $t_0$ is difficult, and given this parameter is co-variant with $\kappa$ and $E_0$ makes analysing all gamma-ray bursts onerous. As mentioned previously, $t_{0}$ can be constrained by identifying the time of a spectral change which marks the transition from the prompt to the afterglow. In practice, this is difficult given the uncertainties on the data. Furthermore, given typical \swift{} slew times, it is often missed entirely. 
This problem of measuring $t_0$ can be alleviated if there are sufficient observations in the transition between the tail of the prompt and the plateau as for gamma-ray bursts analysed here. However, there are notable exceptions, such as GRB130603B which do not have such observations.  
\section{Conclusion} \label{sec:conclusion}
We have introduced a new model for the X-ray afterglow incorporating radiative losses at the shock interface with spin down of a magnetar central engine through an arbitrary braking index. 
By including this new model with emission from the tail of the prompt, we find we can naturally explain a variety of X-ray flares that produce an excess at the onset of the plateau phase. We find that radiative loss can explain both the diversity and sizes of such X-ray flares. In our model, these flares are the result of an energy breakout. 

We also fit our model to a small subset of long and short gamma-ray bursts, the sample selected as they have extensive observations and have been previously suggested to have millisecond magnetar central engines.
In the process, we measure the braking index of eight putative magnetars born in gamma-ray bursts. 
We find these braking indices to be lower than other works \citep[e.g.,][]{xiao19, lu19}, which did not take into account radiative losses and assumed that the X-ray luminosity is only generated through vacuum dipole radiation.
We perform Bayesian model selection between our newly-derived model and one that does not take into account radiative losses, finding for all gamma-ray bursts analysed radiative losses can better explain the data. 

We find that radiative loss can naturally explain the diversity of X-ray plateaus by altering the radiative efficiency $\kappa$ which is a function of the hydrodynamical evolution of the shock and the fraction of total energy transferred to electrons.
However, probing this further requires jointly fitting different electromagnetic bands with X-rays or developing a model that incorporates both the emission from the external shock and the emission from the central engine.
We leave this extension, the exploration of the radiative efficiency, and application of this model to a larger catalogue of short and long gamma-ray bursts to future work. 
\section{Acknowledgments}
We thank the anonymous referee for their comments.
We are grateful to Antonia Rowlinson and Guilia Stratta for helpful discussions. This research was supported by an Australian Government Research Training Program (RTP) Scholarship. P.D.L. is supported through Australian Research Council Future Fellowship FT160100112 and CE170100004. P.D.L and G.A are supported by ARC Discovery Project DP180103155. This work made use of data supplied by the UK Swift Science Data Centre at the University of Leicester.

\section{Data Availability}
All the data used in this paper is available at the UK Swift Science Data Centre at the University of Leicester~\citep{swift_data}.


\bibliographystyle{mnras}
\bibliography{ref}


\bsp    
\label{lastpage}
\end{document}